# PRESERVATION OF THE HIGH QUALITY FACTOR AND ACCELERATING GRADIENT OF Nb$_3$Sn-COATED CAVITY DURING PAIR ASSEMBLY

G. Eremeev*[1], U. Pudasaini[2], S. Cheban[1], J. Fischer[2], D. Forehand[2],
S. Posen[1], A. Reilly[2], R. Rimmer[2], B. Tennis[1]
[1]Fermi National Accelerator Laboratory, Illinois, USA
[2]Thomas Jefferson National Accelerator Facility, Virginia, USA

*Abstract*

Two CEBAF 5-cell accelerator cavities have been coated with Nb$_3$Sn film using the vapor diffusion technique. One cavity was coated in the Jefferson Lab Nb$_3$Sn cavity coating system, and the other in the Fermilab Nb$_3$Sn coating system. Both cavities were measured at 4 K and 2 K in the vertical dewar test in each lab and then assembled into a cavity pair at Jefferson Lab. Previous attempts to assemble Nb$_3$Sn cavities into a cavity pair degraded the superconducting properties of Nb$_3$Sn-coated cavities. This contribution discusses the efforts to identify and mitigate the pair assembly challenges and will present the results of the vertical tests before and after pair assembly. Notably, one of the cavities reached the highest gradient above 80 mT in the vertical test after the pair assembly.

## INTRODUCTION

As the part of the development of Nb$_3$Sn for SRF applications, in 2018 two Nb$_3$Sn-coated CEBAF cavities were assembled into a cavity pair, the standard step during CEBAF cryomodule assembly process. As a part of the pair qualification process, both coated cavities assembled into the cavity pair were measured in the vertical dewar and were found to degrade significantly from their pre-pair assembly qualification tests, Fig.1. The pair was taken apart and each cavity was measured in the vertical dewar separately. These tests confirmed that superconducting RF properties of both Nb$_3$Sn-coated cavities degraded. [1].

Subsequent studies revealed that Nb$_3$Sn-coated SRF cavities are unexpectedly very sensitive to room temperature mechanical tuning, which is a part of the standard process to prepare SRF cavity for cryomodule integration. In the vertical dewar tests of Nb$_3$Sn-coated cavities the low-field surface resistance was observed to increase by about 100 n$\Omega$ and exhibit strong field dependence after room temperature mechanical tuning of a few hundred kilohertz. In a different study by Posen et al., it was found that mechanical tuning done at cryogenic temperatures does not impact coated cavity performance significantly [2].

In order to mitigate the performance degradation after the room temperature tuning, the mechanical tuning step was eliminated for the next Nb$_3$Sn-coated cavity pair assembly. In addition to eliminating room temperature tuning, several improvement were made to the cavity preparation process, e.g., special procedure was developed to shield the inner surface of cavities during chemical etching of niobium flanges in order to reduce the exposure of the coated inner surface to BCP vapor during treatment. Two new CEBAF 5-cell cavities of C75 shape [3] were procured from a commercial vendor, baselined in the vertical dewar test, coated with Nb$_3$Sn, and qualified for the pair assembly at Jefferson Lab.

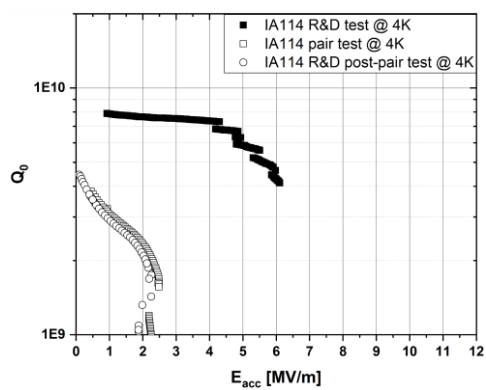

Figure 1: Nb$_3$Sn-coated cavity performance before and after the first Nb$_3$-coated cavity pair assembly. Note the increase in the low-field surface resistance and strong field dependence in the pair test and the test after the pair assembly.

With the adopted changes in the assembly process the coated cavity were assembled into the cavity pair and tested in the vertical dewar. The low-field performance of the coated cavities were largely preserved as compared to the qualification test: one cavity did not exhibited any degradation up to $E_{acc}$ = 3 MV/m, Fig.2, and the low-field surface resistance in the other cavity increased by about 10 n$\Omega$. However, both cavities showed strong Q-slope above few MV/m of accelerating gradient, which limited field reach to below $E_{acc}$ = 10 MV/m in both cavities.

The strong Q-slope degradation showed similarity in both cavities after the pair assembly. It was noticed that the resonant frequency of one of the cavities shifted by about 300 kHz from its value before the pair assembly. Although the other cavity showed little shift in the fundamental frequency, we decided to re-evaluate all the assembly steps for mechanical stress again. Since these cavities are treated above 1100 °C during coating, they are significantly softer than the typical niobium cavity and it was important to assess which steps may cause plastic changes during cavity preparation and pair assembly.

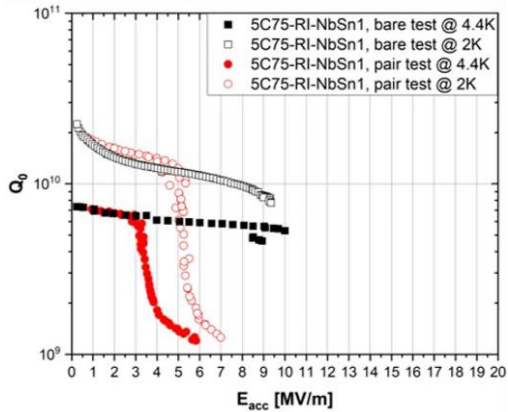

Figure 2: Change in Nb$_3$Sn-coated cavity performance after the second pair assembly with the new mitigation measures implemented. Note that the low-field surface resistance was preserved, but strong Q-slope limits the gradient.

This contribution discusses mechanical analysis to understand the potential causes of frequency shits, the mitigation measures, and the results of the Nb$_3$Sn-coated cavity performance in the cryogenic dewar testing after another pair assembly.

## MECHANICAL SIMULATIONS

To better understand the cause for degradation and the frequency shifts after pair assemblies, mechanical simulations have been done to assess stresses on the cavities during different stages of string assembly. While we checked the stresses on the cavity in different configurations, including, for example, HPR fixtures, most focus was devoted to the handling experienced by cavities during pair assembly. In Fig.3, the cavity pair on the support fixture called strongback is shown. The cavities are mounted onto the strongback after the final HPR. In this fixture each cavity is supported by two mechanical supports at each end. The supports allow the cavities to slide freely along the axial direction, but constrain their radial motion. With mechanical simulation we looked at how support points affect distribution of stresses in the cavity. As it can be expected for this configuration, the stress were found to be well below plastic limits for niobium even after 1200 °C treatment in this configuration.

While on the strongback, the two cavities are assembled with HOM and FPC waveguides, the end dish and joined together by the inner adapter. The cavities than form one hermetic pair that is pumped down and leak checked. The next critical step is the testing of the pair in cryogenic dewar at the vertical test facility. In order to install the pair into the vertical dewar, the strongback with the pair is turned vertical and the pair is transferred from the pair to the vertical attachment on the vertical test stand, Fig.4. The vertical attachment has recently been modified to constrain the pair better. The analysis of the cavity transfer and the cavity hanging in the vertical attachment on the test stand did not indicate any evidence for plastic limits to be exceeded under normal conditions. After the pair is mounted to the vertical attachment on the test stand, it is leak checked and is moved out of the cleanroom with the overhead crane to the cryogenic dewar. This cryogenic test of the cavity pair serves two purposes. One goal is to confirm that the pair was cleanly assembled and meets field emission specifications for the project. The other goal is to verify that the pair is hermetic and maintains its vacuum integrity in the superfluid liquid helium bath: the condition identical to what is seen by the pair inside the helium vessel during operation.

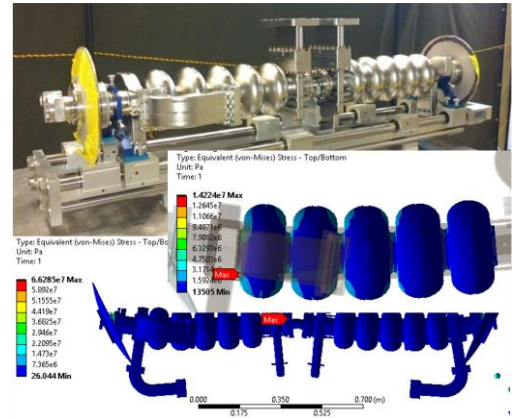

Figure 3: [top]Nb$_3$Sn-coated cavity pair assembled on the strongback. Each cavity is supported from the bottom at each end with three-point fixture.[center] Close-up on the stress distribution in the cavity. [bottom] Stress distribution in the cavity on the strongback in horizontal position. Note that the stresses are well below the plastic limit.

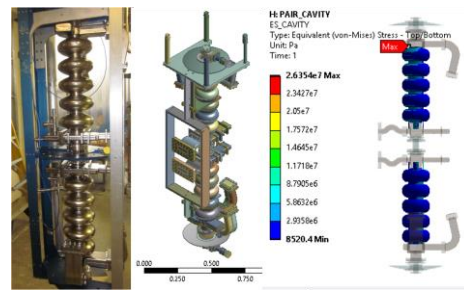

Figure 4: [left]Nb$_3$Sn-coated cavity pair assembled on the vertical test stand in preparation for cryogenic testing in the vertical dewar. [center] CAD drawing of the modified vertical attachment support fixture [bottom] Stress distribution in the cavity on the vertical test stand. Note that the stresses are higher than in the horizontal position, but are still below the plastic limit.

During the analysis of the mechanical simulation results, we realized that, while cavities are well supported and cavity handling with the manual lifts should not induce stresses in excess of niobium material limits, the crane movement of the pair is harder to constraint and control and special damping fixtures need to be designed and tested to reduce the risk of mechanical deformation during this step. The design and testing of such fixture was beyond the scope of this project, so, to mitigate the potential issue with crane moves, the decision was made to skip cryogenic dewar test of the cavity pair and proceed directly to the cryomodule assembly.

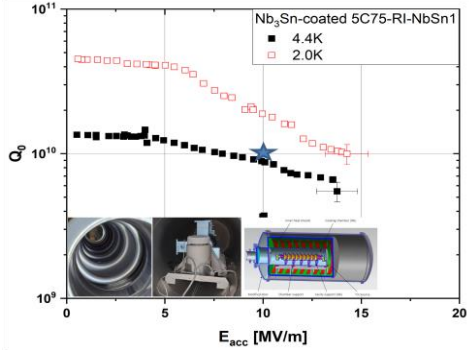

Figure 5: Qualification test results of the cavity coated at Fermilab Nb$_3$Sn-coating facility. Note that the cavity gradient exceeds E$_{acc}$ = 14 MV/m and Q$_0$ reaches $10^{10}$ at E$_{acc}$ = 10 MV/m at 4.4 K, which is the goal. [inset] Pictures of the cavity assembled for the coating in the furnace, the picture of the inside surface of coated cavity, and the coating furnace layout reproduced from [5].

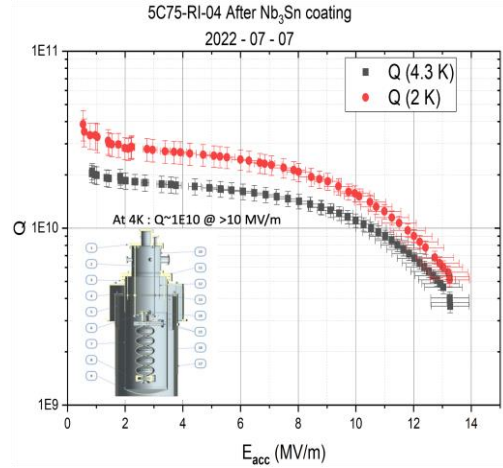

Figure 6: Qualification test results of the cavity coated at JLab Nb$_3$Sn-coating facility. Note that the cavity gradient exceeds E$_{acc}$ = 13 MV/m and Q$_0$ exceeds $10^{10}$ at E$_{acc}$ = 10 MV/m at 4.4 K, which is the specification. [inset] the coating furnace layout reproduced from [4].

## PAIR PERFORMANCE PRESERVATION AFTER PAIR ASSEMBLY

For the next pair assembly, another round of re-processing and re-coating with Nb$_3$Sn of two cavities was initiated. The decision was made to coat one cavity at Fermilab Nb$_3$Sn coating facility and the other cavity at Jefferson Lab Nb$_3$Sn coating facility with the goal to expedite cavity re-qualification process. CEBAF 5-cell cavity was processed and coated at Fermilab Nb$_3$Sn coating facility for the first time. During the re-processing, the other cavity developed a leak in the weld joint of the fundamental power coupler waveguide. The reason for the leak turned out to be the thinning on one of the welds in the end groups of the cavity. While the cells in this cavity were made out of new niobium sheets, the end groups were taken from one of the old cavities from the original CEBAF cavity production by Interatom, in order to save on the cost of the cavity fabrication. One of these end groups, thinned by the present and previous chemical treatments, developed a leak after additional processing. Few attempts to fix the leak by electron beam welding proved unsuccessful. The cavity was thus unsuitable for additional coatings and had to be replaced with another niobium cavity. The newly selected cavity was also C75 shape, but, unlike other cavities discussed in this contribution, made out of large grain niobium material. The efforts to re-coat this cavity progressed smoothly and the cavity was successfuly coated with Nb$_3$Sn at Jefferson Lab Nb$_3$Sn coating facility. In Fig. 5 and Fig. 6 the vertical test results after the latest Nb$_3$Sn coatings of CEBAF 5-cell cavities from both facilities are shown. The vertical test performance of both cavities exceeded the gradient specification of 10 MV/m and met or exceeded the quality factor specification of $10^{10}$ at 4 K.

After the vertical cavity qualification tests, both cavities were prepared and assembled into the cavity pair. During pumpdown of the cavity pair after the pair assembly, a warm leak was discovered in the ceramic window of one of the fundamental power coupler waveguides. The pair had to be disassembled to replace the leaking waveguide. We decided to use this opportunity to take the pair completely apart and to test both cavities in the vertical dewar again to assess the impact of additional measures in helping to preserve the performance of Nb$_3$Sn-coated cavities.

Both cavities were completely disassembled, cleaned, assembled with vertical test hardware and tested in the vertical dewars. One of the cavities exhibited degradation in the vertical dewar test, Fig.7, but the improvement in the low-field quality factor and the reduction in the Q-slope degradation were observed as compared to tests after the previous pair assembly attempt, Fig.2.

The second cavity did not show any noticeable degradation. In the vertical test at 2 K after the pair assembly, the cavity reached E$_{acc}$ ≃ 20 MV/m, corresponding to 80 mT of the peak surface magnetic field, after multipacting barrier was processed at around E$_{acc}$ = 15 MV/m. This is the best performance and the highest accelerating gradient reached in Nb$_3$Sn-coated 5-cell cavities.

Since the total energy gain of two cavities still exceeds 10 MeV goal and one of the cavities preserved high quality factor, we are progressing these cavities to the cavity pair assembly without any additional re-processing and re-coating. As of this writing the cavity pair has been assembled, passed the leak check, and is being integrated into the helium vessel.

## CONCLUSION

We investigated the degradation of Nb$_3$Sn-coated cavities after preparation and assembly into the cavity pair. CEBAF

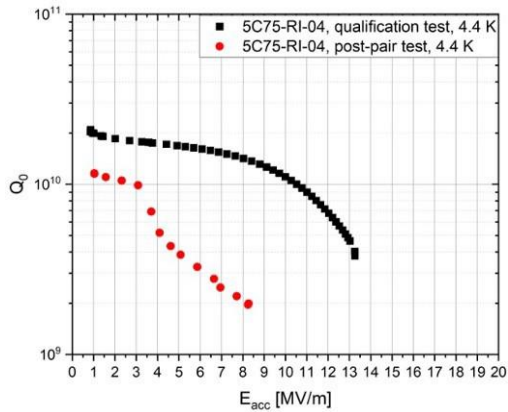

Figure 7: Nb$_3$Sn-coated cavity performance before and after the pair assembly with the new mitigation measures implemented. Note that the low-field surface resistance was preserved, but strong Q-slope limits the gradient.

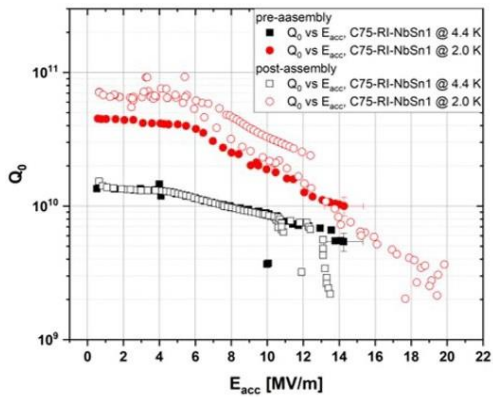

Figure 8: Nb$_3$Sn-coated cavity performance before and after the pair assembly with the new mitigation measures implemented. No change in the cavity performance was observed. After the pair assembly, the cavity reached close to 20 MV/m accelerating gradient, which corresponds to about 80 mT of the peak surface magnetic field.

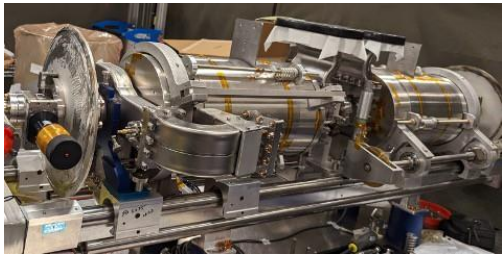

Figure 9: Nb$_3$Sn cavity pair prepared for helium vessel integration. The original CEBAF cavities did not have local magnetic shielding only global magnetic shielding. This pair, similar to other C75 pair, employs both local and global magnetic shielding [6].

cavity pair asembly is the milestone for integrating cavities into CEBAF cryomodule. In the previous attempts, significant degradation in cavity performance was observed after pair assembly. Several causes, such as warm tuning, were identified and mitigated to maintain the quality factor and gradient reach from vertical dewar qualification test to pair assembly. As an additional precaution, cavity pair test in the vertical dewar after pair assembly, which is the standard step in CEBAF cryomodule assembly, was eliminating from the preparation process. As the result of these measures, one cavity exhibited some degradation, while the other cavity fully maintained its performance after the pair assembly. The best cavity reached close to 20 MV/m accelerating gradient, which corresponds to about 80 mT of the peak surface magnetic field. Since the total energy gain of the pair still exceeds 10 MeV, the cavities are progress to cryomodule assembly without reprocessing and re-coating.

## ACKNOWLEDGMENTS

This manuscript has been authored by Fermi Research Alliance, LLC under Contract No. DE-AC02-07CH11359 with the U.S. Department of Energy, Office of Science, Office of High Energy Physics. This material is based upon work supported by the U.S. Department of Energy, Office of Science, Office of Nuclear Physics.